\documentclass[journal]{IEEEtran}
\usepackage[T1]{fontenc}
\usepackage{graphicx}
\usepackage{booktabs}
\usepackage{amsmath,amssymb}
\usepackage{array}
\usepackage{multirow}
\usepackage{url}
\usepackage{cite}
\usepackage{stfloats}
\usepackage{placeins}
\usepackage[svgnames]{xcolor}
\usepackage[hidelinks]{hyperref}
\usepackage{orcidlink}
\graphicspath{{Figs/}}
\newcommand{\ignore}[1]{}
\usepackage[table]{xcolor}

\usepackage{tikz}
\usetikzlibrary{arrows.meta, positioning}
\usetikzlibrary{patterns, arrows.meta}

\begin{document}

\bstctlcite{IEEEexample:BSTcontrol}

\title{A Low-Latency ASIC Architecture for Real-Time Line Segment Detection}

\author{
Amir Hossein Jalilvand$^\pi$~\orcidlink{0000-0002-7641-6606},
Parsa Hassani Shariat Panahi$^\pi$~\orcidlink{0009-0005-2912-3754},
and M. Hassan Najafi$^+$~\orcidlink{0000-0002-4655-6229} \\
\thanks{$^\pi$School of Computer Engineering, Iran University of Science and Technology, Tehran, Iran\\
$^+$Electrical, Computer, and Systems Engineering Department, Case Western Reserve University, OH, USA}
}

\maketitle


\begin{abstract}
Line segment detection is a critical preprocessing step in embedded vision applications such as autonomous navigation, visual SLAM, and industrial inspection. Deep learning methods achieve high accuracy but require substantial resources, limiting their deployment on resource-constrained platforms. Classical algorithms are efficient but exhibit content-dependent latency. This paper presents a low-latency ASIC architecture for real-time line segment detection. The proposed design is based on the step-length algorithm and incorporates five ASIC-specific features: register-based line buffering with data reuse, multiplierless MCM-based filtering, 8-class angle quantization, a CAM-like associative memory for single-cycle matching, and an optimized duplicate removal mechanism. The architecture is fully pipelined and processes one pixel per clock cycle with deterministic latency. Synthesized in a 45nm CMOS process, the design achieves 325 FPS at VGA resolution and 48 FPS at Full HD, with 25.54 mW power consumption and 0.412 mm\textsuperscript{2} area. At 125 MHz, the throughput increases to 406 FPS at VGA resolution with 31.48 mW power consumption. Compared with a 90nm ASIC implementation based on the Line Hough Transform, the proposed design reduces power consumption by 49\% and delivers over 1.6 times higher frame rate. The architecture is well suited for edge-computing applications requiring real-time performance, low power, and minimal area.
\end{abstract}


\section{Introduction}\label{sec:intro}
\IEEEPARstart{L}{ine} segment detection is a fundamental preprocessing step in computer vision, serving as a critical enabler for applications ranging from autonomous navigation and robotics to augmented reality and industrial inspection~\cite{lin2024comprehensive, vongioi2010lsd, akinlar2011edlines}. Accurate and real-time detection of line segments is essential for visual SLAM, vanishing-point estimation, lane detection, power-line monitoring, and 3D reconstruction~\cite{hu2025emlsd}. Despite extensive research, the field continues to face challenges in robustness, efficiency, and hardware deployment, particularly for resource-constrained embedded systems~\cite{lin2024comprehensive}.

Classical line segment detectors such as LSD~\cite{vongioi2010lsd} and EDLines~\cite{akinlar2011edlines} are designed for general-purpose CPUs and rely on hand-crafted gradient-based pipelines. While these methods are lightweight compared to deep learning approaches, their runtime depends on image content, and their accuracy degrades under blur, low contrast, and clutter. ELSED~\cite{suarez2021elsed} improves speed on embedded CPUs but remains fundamentally a classical detector with no learning-based adaptation. Recent learned wireframe parsers including L-CNN~\cite{zhou2019lcnn}, HAWP~\cite{xue2020hawp}, LETR~\cite{xu2021letr}, and ULSD~\cite{li2021ulsd} achieve high accuracy but require GPU-class compute and several megabytes of memory. Lightweight variants such as M-LSD~\cite{gu2022mlsd}, LSDNet~\cite{teplyakov2022lsdnet}, and EM-LSD~\cite{em-lsd} target mobile platforms, yet still exceed the memory and power budgets of low-cost microcontrollers and high-volume ASIC implementations. Even the most memory-efficient learned detector~\cite{MiLSD2026} operates on microcontrollers using quantized neural networks, but relies on software inference rather than custom hardware acceleration, leaving the ASIC design space largely unexplored.

To overcome the limitations of software-based detectors, several hardware implementations have been proposed. The Hough transform has been mapped to ASIC using CORDIC-based architectures~\cite{Majumdar2000ASIC}, with later memory-efficient designs~\cite{Pachkor2018Memory}. More recently, the step-length algorithm was demonstrated on FPGA~\cite{Ossimitz2021FPGA}, showing deterministic real-time latency with low resource utilization. However, existing hardware accelerators face two key limitations: (i) Hough-transform-based ASICs rely on large parameter-space memories and CORDIC computations, which limit scalability and energy efficiency; and (ii) FPGA implementations, while flexible, are not optimal for high-volume, low-cost embedded systems due to the overhead of programmable logic and the absence of BRAM and DSP resources in ASIC designs.

This work introduces a novel ASIC architecture that addresses these limitations by building upon the step-length algorithm and incorporating five ASIC-specific enhancements: (i) register-based line buffering with data reuse, (ii) multiplierless MCM-based filtering, (iii) 8-class angle quantization, (iv) a CAM-like associative memory for single-cycle chain matching, and (v) an optimized sliding-window duplicate removal mechanism. Unlike the original FPGA implementation, our design eliminates all multipliers and BRAMs, enabling deterministic single-cycle matching with significantly lower area and power consumption.

The key contributions of this paper are:

\begin{itemize}
    \item \textbf{Register-based line buffering with data reuse:} Replaces BRAM-based storage with a register array and circular pointer mechanism, reducing switching activity and eliminating memory controllers.
    \item \textbf{Multiplierless MCM-based filtering:} Replaces all multipliers in the Gaussian and Sobel filters with shift-and-add operations, eliminating 27 multipliers per pixel.
    \item \textbf{8-class angle quantization:} Extends the baseline 4-class quantization to 8 classes (3 bits), improving diagonal line detection accuracy with minimal area overhead.
    \item \textbf{CAM-like associative chain builder:} Enables single-cycle chain matching with a fixed capacity of 64 active chains, reducing storage requirements by 95\% compared to conventional FIFO-based approaches.
    \item \textbf{Optimized sliding-window duplicate removal:} Uses angular pre-filtering (XOR-based) to reduce switching activity before full geometric comparison.
\end{itemize}

The proposed design is fully pipelined, processing one pixel per clock cycle with deterministic latency independent of scene content. Synthesized in a 45nm CMOS process, it achieves 325 FPS at VGA resolution and 48 FPS at Full HD, with only 25.54 mW power consumption and 0.412 mm\textsuperscript{2} area. At 125 MHz, the design achieves 406 FPS at VGA resolution with 31.48 mW power consumption. The architecture is well-suited for edge-computing applications where real-time performance, low power, and minimal silicon area are critical.

The remainder of this paper is organized as follows. Section~\ref{sec:asic_architecture} presents the proposed ASIC architecture in detail, covering both the pixel-level front-end and the segment-level back-end. Section~\ref{sec:implementation_result} presents the RTL implementation, synthesis results, and performance comparisons. Finally, Section~\ref{sec:conclusion} concludes the paper.

\section{Proposed Architecture of the Line Segment Detector}
\label{sec:asic_architecture}

\begin{figure}[]
    \centering
    \includegraphics[
        width=0.75\columnwidth,
        trim=0.5cm 0.5cm 0.5cm 0.5cm,
        clip
    ]{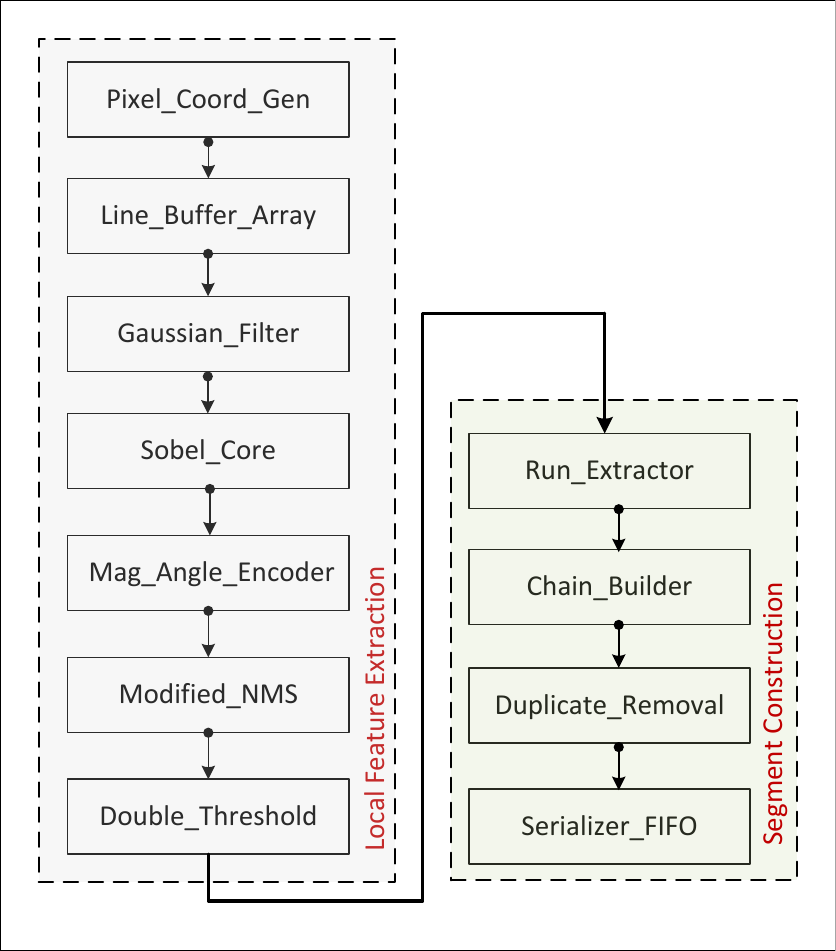}
   \caption{Architecture of the Line Segment Detection Pipeline, illustrating the flow from pixel processing to segment extraction.}
    \label{fig:lsd_diagram_detailed}
\end{figure}

This section presents the ASIC architecture of the proposed line segment detector, optimized for standard-cell CMOS implementation. The architecture implements the step-length algorithm with three key features in the front-end: (i) register-based line buffering with data reuse, (ii) multiplierless Multiple Constant Multiplication (MCM)-based filtering, and (iii) 8-class angle quantization. For the back-end, it employs (iv) a CAM-like associative memory enabling single-cycle chain matching, and (v) an optimized sliding-window duplicate removal mechanism. The design is fully pipelined, processing one pixel per clock cycle with deterministic latency independent of scene content.
The pipeline is divided into two synergistic stages: the \textit{Pixel-Level Streaming Front-End} and the \textit{Segment-Level Back-End}. This division ensures modularity, scalability, and adherence to strict real-time processing constraints. The overall system is depicted in Fig.~\ref{fig:lsd_diagram_detailed}.

\subsection{Pixel-Level Streaming Front-End: Edge Feature Extraction}
\label{subsec:frontend_asic}
The front-end stage transforms the incoming 8-bit grayscale pixel stream into a binary edge map with 3-bit angle class information. It follows the conventional Canny-based edge detection pipeline, which includes Gaussian smoothing, Sobel gradient calculation, non-maximum suppression, and double thresholding. Three features are specifically tailored for ASIC implementation: register-based line buffering with data reuse, multiplierless MCM-based filtering, and 8-class angle quantization.

The pipeline starts with the \texttt{Pixel\_Coord\_Gen} block. This unit synchronizes the incoming pixel stream, \texttt{pixel\_in[7:0]}, with its associated validity signal, \texttt{pixel\_valid}. It also generates the pixel coordinates within the image frame, denoted as \texttt{x\_coord[W-1:0]} and \texttt{y\_coord[H-1:0]}, where \(W\) and \(H\) are the image width and height, respectively. The implementation uses combinational logic for coordinate calculation, typically based on incrementing counters, and propagates the validity signal (\texttt{pix\_valid\_o}) through the pipeline.

\subsubsection{Register-Based Line Buffering with Data Reuse}
\label{subsubsec:line_buffer}

To enable neighborhood operations, the front-end stores three rows of pixel data, which allows single-cycle access to a \(3 \times 3\) window centered on the current pixel. This is achieved using a register-based storage array with a pointer-based circular shift mechanism. Only the newly arrived pixel is written into the buffer each cycle; the remaining window pixels are accessed through multiplexed read ports. This data reuse technique ensures deterministic access latency and removes the need for dedicated memory controllers. The resulting \(3 \times 3\) window (\texttt{win\_p00..p22[7:0]}) is then provided to subsequent processing blocks.

Table~\ref{tab:line_buffer_example} illustrates a concrete example of this process. Assume an image width of \(W = 10\), where the buffer has been pre-filled with the values shown in the table. The circular pointer \texttt{ptr} is set to \(5\), indicating that the next incoming pixel will be written at index \(5\) in Row 0.

\begin{table}[htbp]
\centering
\caption{Example of the \(3 \times 3\) window extracted from the line buffer when \(ptr = 5\).}
\label{tab:line_buffer_example}
\begin{tabular}{c|cccccccccc}
\hline
\textbf{Index} & 0 & 1 & 2 & 3 & 4 & 5 & 6 & 7 & 8 & 9 \\
\hline
\textbf{Row 0} & 10 & 11 & 12 & \cellcolor{gray!20}13 & \cellcolor{gray!20}14 & \cellcolor{gray!20}15 & 16 & 17 & 18 & 19 \\
\textbf{Row 1} & 20 & 21 & 22 & \cellcolor{gray!20}23 & \cellcolor{gray!20}24 & \cellcolor{gray!20}25 & 26 & 27 & 28 & 29 \\
\textbf{Row 2} & 30 & 31 & 32 & \cellcolor{gray!20}33 & \cellcolor{gray!20}34 & \cellcolor{gray!20}35 & 36 & 37 & 38 & 39 \\
\hline
\end{tabular}
\end{table}

When a new pixel arrives, it is written into Row 0 at the address indicated by \texttt{ptr}. The \(3 \times 3\) window is then formed by reading indices \(\texttt{ptr}-2\), \(\texttt{ptr}-1\), and \(\texttt{ptr}\) from all three rows. In the next cycle, \texttt{ptr} is updated, and the window shifts accordingly. No data shifting occurs within the buffer; only the pointer changes, while the stored pixel values remain unchanged. This approach reduces switching activity and dynamic power consumption by minimizing the number of register updates per cycle. Furthermore, the window is accessed via combinational multiplexers, which eliminates the need for additional clock cycles and ensures deterministic access latency.

\subsubsection{Multiplierless MCM-Based Filtering}
\label{subsubsec:mcm_filtering}

The \texttt{Gaussian\_Filter} applies a \(3 \times 3\) smoothing kernel to reduce noise and minor intensity variations, thereby improving edge detection robustness. The Gaussian kernel is defined as:

\begin{equation}
G = \frac{1}{16}
\begin{bmatrix}
1 & 2 & 1 \\
2 & 4 & 2 \\
1 & 2 & 1
\end{bmatrix}
\label{eq:gaussian_kernel}
\end{equation}

Instead of using multipliers, the architecture employs Multiple Constant Multiplication (MCM) with shift-and-add operations. Multiplication by 1, 2, and 4 is implemented as no operation, a left shift by one, and a left shift by two, respectively. The complete Gaussian output is computed as:

\begin{equation}
\begin{aligned}
G_{out} = & \ (I_{TL} + I_{TR} + I_{BL} + I_{BR}) \gg 4 \; + \\
         & \ ((I_T + I_L + I_R + I_B) \ll 1) \gg 4 \; + \\
         & \ (I_C \ll 2) \gg 4
\end{aligned}
\label{eq:gaussian_mcm}
\end{equation}

where \(I_{TL}, I_T, \ldots, I_C\) are the \(3 \times 3\) neighborhood pixels, \(\ll\) denotes left shift, and \(\gg 4\) represents division by 16. The output (\texttt{gauss\_out[9:0]}) uses 9-bit precision to prevent overflow during summation.

Next, the \texttt{Sobel\_Core} computes the horizontal (\(G_x\)) and vertical (\(G_y\)) gradient components using the Sobel kernels:

\begin{equation}
K_x = 
\begin{bmatrix}
-1 & 0 & 1 \\
-2 & 0 & 2 \\
-1 & 0 & 1
\end{bmatrix}
\label{eq:sobel_x}
\end{equation}

\begin{equation}
K_y = 
\begin{bmatrix}
-1 & -2 & -1 \\
0 & 0 & 0 \\
1 & 2 & 1
\end{bmatrix}
\label{eq:sobel_y}
\end{equation}

The Sobel operations also employ MCM, where coefficients -2, -1, 0, 1, and 2 are implemented using shifts and negations. The gradient components are computed as:

\begin{equation}
G_x = (I_{TR} - I_{TL}) + ((I_{CR} - I_{CL}) \ll 1) + (I_{BR} - I_{BL})
\label{eq:sobelx_mcm}
\end{equation}

\begin{equation}
G_y = (I_{BL} - I_{TL}) + ((I_{BC} - I_{TC}) \ll 1) + (I_{BR} - I_{TR})
\label{eq:sobely_mcm}
\end{equation}

The gradient components are represented as 9-bit signed values in two's complement form, while the magnitude output is clipped to 8-bit unsigned.

\subsubsection{Magnitude and Angle Encoding}
\label{subsubsec:mag_angle}

The \texttt{Mag\_Angle\_Encoder} block processes the gradient components to approximate the edge magnitude and quantize the gradient direction. The magnitude is computed using the L1 norm approximation, which avoids square root operations:

\begin{equation}
\text{mag} = |G_x| + |G_y|
\label{eq:magnitude_l1}
\end{equation}

This computation is performed using absolute value circuits and adders. The gradient angle is computed as:

\begin{equation}
\theta = \operatorname{atan2}(G_y, G_x)
\label{eq:gradient_angle}
\end{equation}

\begin{figure}
\centering
\begin{tikzpicture}[scale=1]
    \filldraw[fill=gray!40, draw=black, thick] 
        (0,0) -- (-11.25:2.5cm) arc (-11.25:11.25:2.5cm) -- cycle;
    \filldraw[fill=gray!40, draw=black, thick] 
        (0,0) -- (168.75:2.5cm) arc (168.75:191.25:2.5cm) -- cycle;
    
    \foreach \center in {22.5,45,67.5,90,112.5,135,157.5} {
        \filldraw[fill=white, draw=black, thick] 
            (0,0) -- (\center-11.25:2.5cm) arc (\center-11.25:\center+11.25:2.5cm) -- cycle;
        \filldraw[fill=white, draw=black, thick] 
            (0,0) -- (\center+180-11.25:2.5cm) arc (\center+180-11.25:\center+180+11.25:2.5cm) -- cycle;
    }
    
    \foreach \angle in {-11.25,11.25,33.75,56.25,78.75,101.25,123.75,146.25,168.75,191.25,213.75,236.25,258.75,281.25,303.75,326.25} {
        \draw[thick] (0,0) -- (\angle:2.5cm);
    }
    
    \foreach \angle/\class in {0/0,22.5/1,45/2,67.5/3,90/4,112.5/5,135/6,157.5/7} {
        \node at (\angle:1.8cm) {\class};
        \node at (\angle+180:1.8cm) {\class};
    }
    
    \draw[<->, thick, red] (-11.25:2.8cm) arc (-11.25:11.25:2.8cm);
    \node[red, font=\tiny] at (0:3.3cm) {$22.5^\circ$};
    
    \draw[<->, thick, red] (168.75:2.8cm) arc (168.75:191.25:2.8cm);
    \node[red, font=\tiny] at (180:3.3cm) {$22.5^\circ$};
    
    \filldraw (0,0) circle (0.05cm);
\end{tikzpicture}
\caption{8-class angle quantization. Class 0 (gray) is the reference, spanning \(-11.25^\circ\) to \(+11.25^\circ\) and \(168.75^\circ\) to \(191.25^\circ\). Each class spans \(22.5^\circ\).}
\label{fig:angle_quantization}
\end{figure}
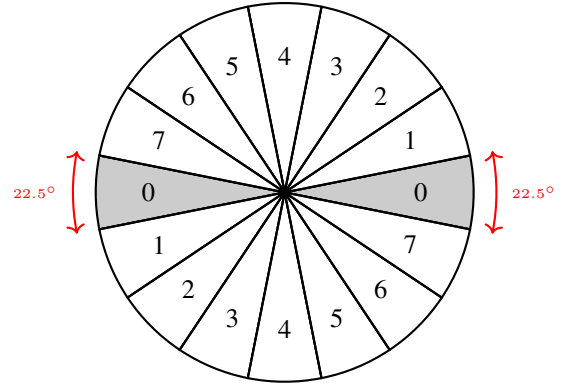

Angle quantization classifies the gradient direction into discrete sectors. The architecture uses 8-class quantization (3 bits), which provides finer angular resolution than 4-class schemes. Fig. ~\ref{fig:angle_quantization} illustrates the 8-class quantization scheme, where each class spans \(22.5^\circ\) and appears in both upper and lower half-planes. Class 0, shown in gray, serves as the reference sector and is centered at \(0^\circ\) and \(180^\circ\).

This quantization requires only 3 bits per pixel for angle information and improves detection accuracy for diagonal line segments compared to 4-class (2-bit) schemes. The angle class is determined by the sector in which the computed angle \(\theta\) falls.


\subsubsection{Modified Non-Maximum Suppression}
\label{subsubsec:modified_nms}

The \texttt{Modified\_NMS} block performs non-maximum suppression to thin detected edges to single-pixel width. In contrast to the standard Canny NMS, which compares only the center pixel with its orthogonal neighbors, the modified algorithm also checks the center pixel's neighbors against their respective orthogonal neighbors. The suppression condition is defined as:

\begin{equation}
\text{edge} = 
\begin{cases}
1, & \text{if } \text{mag}(x,y) \geq \text{mag}(x + \Delta x, y + \Delta y) \\
  & \text{and } \text{mag}(x,y) \geq \text{mag}(x - \Delta x, y - \Delta y) \\
0, & \text{otherwise}
\end{cases}
\label{eq:nms_condition}
\end{equation}

where \((\Delta x, \Delta y)\) is perpendicular to the gradient direction. If the condition is not satisfied, the magnitude is suppressed to zero, which effectively thins the edge. This modified approach improves edge continuity by smoothing over outlier pixels that would otherwise be discarded.

\subsubsection{Double Thresholding}
\label{subsubsec:double_threshold}

The final stage of the front-end is the \texttt{Double\_Threshold} block, which applies hysteresis thresholding to generate the final binary edge map (\texttt{edge\_bit}). The thresholding logic uses fixed thresholds \(T_{\text{low}} = 8\) and \(T_{\text{high}} = 12\) for gradient magnitudes in the range [0, 255]:

\begin{align}
\text{edge\_bit} = 
\begin{cases}
1, & \text{if } \text{mag} \geq 12 \text{ (strong edge)} \\[4pt]
1, & \text{if } 8 \leq \text{mag} < 12 \\ 
  & \text{and connected to strong edge} \\[4pt]
0, & \text{otherwise}
\end{cases}
\label{eq:threshold_fixed}
\end{align}

Pixels with magnitude above \(T_{\text{high}}\) are marked as strong edges. Pixels with magnitude between \(T_{\text{low}}\) and \(T_{\text{high}}\) are considered weak edges and are retained only if connected to a strong edge, either directly or through other weak edges. Pixels below \(T_{\text{low}}\) are discarded. This hysteresis mechanism improves contour continuity while suppressing noise-induced edges. Fixed thresholds eliminate the need for programmable registers, which reduces area and simplifies the control logic.

The output of the front-end is a binary edge map in which each pixel indicates the presence or absence of an edge, along with the associated 3-bit angle class for edge pixels. This data stream is passed to the back-end for line segment extraction.

\subsection{Segment-Level Back-End: From Edges to Lines}
\label{subsec:backend_asic}

The back-end stage processes the binary edge stream generated by the front-end to extract complete line segments. It consists of four cascaded blocks: run extraction, chain building, duplicate removal, and serialization. Unlike the combinational front-end, this stage relies on stateful logic to manage active segments across multiple scanlines.

\subsubsection{Run Extractor}
\label{subsubsec:run_extractor}

The \texttt{Run\_Extractor} is the first stage of the back-end. It converts the binary edge stream into a sequence of run-length encoded primitives. The block scans the incoming binary edge map (\texttt{edge\_bit}) on a per-scanline basis and detects contiguous horizontal runs of edge pixels, which are groups of consecutive edge pixels belonging to the same row.

For each detected run, the block outputs a compact descriptor containing:
\begin{itemize}
    \item The start and end X-coordinates (\texttt{run\_x\_start[W-1:0]}, \texttt{run\_x\_end[W-1:0]})
    \item The Y-coordinate of the scanline (\texttt{run\_y[H-1:0]})
    \item The dominant gradient angle class (\texttt{run\_angle[2:0]}), derived from the majority angle among edge pixels within the run
\end{itemize}

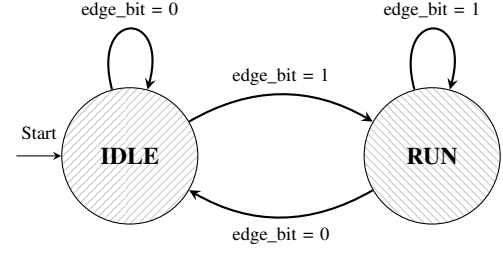
\begin{figure}[]
\centering
\begin{tikzpicture}[
    node distance=4cm,
    state/.style={circle, draw, minimum width=1.8cm, minimum height=1.8cm, align=center, font=\bfseries\small},
    arrow/.style={->, >=stealth, thick}
]

    \node[state, pattern=north east lines, pattern color=black!20] (IDLE) {IDLE};
    \node[state, pattern=north west lines, pattern color=black!20, right of=IDLE] (RUN) {RUN};

    \draw[->, >=stealth] (-1.5,0) -- (IDLE);
    \node[font=\scriptsize] at (-1.2,0.3) {Start};

    \draw[arrow] (IDLE) edge[loop above, looseness=6] node[above, font=\scriptsize] {edge\_bit = 0} (IDLE);
    \draw[arrow] (RUN) edge[loop above, looseness=6] node[above, font=\scriptsize] {edge\_bit = 1} (RUN);

    \draw[arrow] (IDLE) edge[bend left, above] node[above, font=\scriptsize] {edge\_bit = 1} (RUN);
    \draw[arrow] (RUN) edge[bend left, below] node[below, font=\scriptsize] {edge\_bit = 0} (IDLE);

\end{tikzpicture}
\caption{Two-state finite-state machine of the Run Extractor. Transitions are triggered by the \texttt{edge\_bit} signal.}
\label{fig:run_extractor_fsm}
\end{figure}

The implementation uses a simple two-state finite-state machine (FSM) to track edge continuity. Fig. ~\ref{fig:run_extractor_fsm} illustrates the state transitions of the FSM.

In the IDLE state, the FSM waits for the first edge pixel of a run. When \texttt{edge\_bit} transitions to 1, it captures the current X-coordinate as \texttt{run\_x\_start}, initializes the angle histogram, and moves to the RUN state. In the RUN state, the FSM continues scanning while \texttt{edge\_bit} remains 1, updating \texttt{run\_x\_end} and recording the angle class of each edge pixel. When \texttt{edge\_bit} transitions back to 0, the FSM outputs the accumulated run descriptor and returns to the IDLE state.

This block uses only sequential logic and basic comparators. It requires no memory or arithmetic units beyond simple counters. A small output FIFO is employed to absorb the dynamic rate of run generation and to decouple the run extractor from the subsequent chain builder.

\subsubsection{CAM-like Chain Builder}
\label{subsubsec:chain_builder}

The \texttt{Chain\_Builder} is the core processing element of the back-end. It links run segments across adjacent scanlines to form coherent line segments. The architecture employs a semi-associative CAM-like (Content-Addressable Memory) structure with a fixed capacity of 64 active chains. This design choice is motivated by the target edge-computing applications, including autonomous driving, drone navigation, and industrial inspection, where real-time processing, low power consumption, and minimal silicon area are critical. In these domains, the number of simultaneously active line segments rarely exceeds 64. This observation is supported by analysis of standard datasets representative of edge-computing workloads, including YorkUrban~\cite{yorkurban} and TESTIMAGES~\cite{testimages}, where over 95\% of scenes contain fewer than 64 active chains at any given time.

\begin{figure}
\centering
\begin{tikzpicture}[
    node distance=0.7cm and 0.3cm,
    block/.style={rectangle, draw, minimum width=2.5cm, minimum height=0.7cm, align=center, font=\small},
    arrow/.style={->, >=stealth, thick}
]

    \node[block, fill=white] (run) {Run Descriptor};

    \node[block, fill=black!5, below=of run] (cam) {CAM Array (64 Entries)};

    \node[block, pattern=north east lines, pattern color=black!30, below=of cam] (ang) {Angular Filter (XOR)};

    \node[block, pattern=north west lines, pattern color=black!30, below=of ang] (spat) {Spatial Match (Parallel)};

   
    \node[block, pattern=north east lines, pattern color=black!30, below=of spat] (early) {Early Finalization};

    \node[block, fill=white, below=of early] (seg) {Segment Descriptor};

    \draw[arrow] (run) -- (cam);
    \draw[arrow] (cam) -- (ang);
    \draw[arrow] (ang) -- (spat);
    \draw[arrow] (spat) -- (early);
    \draw[arrow] (early) -- (seg);

    \node[left=0.1cm of ang, font=\scriptsize, align=right] {Stage 1};
    \node[left=0.1cm of spat, font=\scriptsize, align=right] {Stage 2};
    \node[left=0.1cm of early, font=\scriptsize, align=right] {Stage 3};

\end{tikzpicture}
\caption{CAM-like Chain Builder architecture with three pipelined stages: angular filtering, parallel spatial matching, and early finalization. Distinct fill patterns distinguish stages for black-and-white printing.}
\label{fig:chain_builder}
\end{figure}
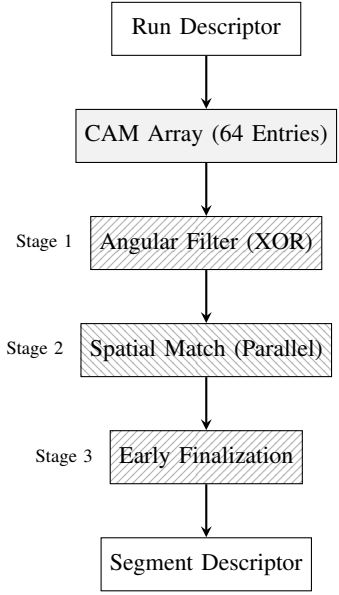

Each entry in the register array stores the metadata of an ongoing line segment. The structure of each chain record is illustrated in Fig. ~\ref{fig:chain_builder}.

\begin{enumerate}
    \item \textbf{Coarse Angular Filtering:} For each incoming run, a lightweight XOR-based comparison is performed against the angle class of all active chains. This step disqualifies any chain whose angle class deviates by more than one class (\textit{i.e.,} \(22.5^\circ\)) from the run's angle, drastically reducing the number of candidates that proceed to the spatial matching stage.
    
    \item \textbf{Parallel Spatial Matching:} The remaining candidates are evaluated in parallel against the current run using dedicated hardware comparators. The matching criteria require that the run resides on the immediately adjacent scanline and that its horizontal coordinate overlaps with the current endpoint of the chain. If a compatible match is found, the chain's endpoint and length are updated accordingly. If no match is found, the incoming run initializes a new active chain.
\end{enumerate}

To illustrate the matching process, consider an active chain with endpoint at \((x=100, y=40)\), angle class 2, and length 25. A new run arrives with \(x_{\text{start}}=98\), \(x_{\text{end}}=102\), \(y=41\), and angle class 2. The angular filter passes the chain (same angle class). The spatial matching logic checks:
\begin{itemize}
    \item Vertical proximity: \(y_{\text{run}} = 41\) and \(y_{\text{chain}} = 40\) (adjacent scanline)
    \item Horizontal overlap: \(x_{\text{start}}=98 \leq 100+1\) and \(x_{\text{end}}=102 \geq 100-1\) (overlap exists)
\end{itemize}
The match is successful, and the chain is updated with the new endpoint \((x=102, y=41)\). The length becomes \(25 + (102-98) + 1 = 30\).

For rare scenarios where the number of active chains exceeds 64, we employ an early finalization strategy: the oldest chains—those that have not been updated for a configurable number of scanlines—are finalized and output as complete segments. For example, if the oldest chain has endpoints \((10, 5)\) and \((20, 15)\), it is output as a segment with length:
\[
\text{seg\_len} = |20-10| + |15-5| = 20
\]
and its angle class is preserved in the segment descriptor.
This early finalization ensures that the register array never becomes full while preserving detection integrity. The CAM-like structure enables deterministic single-cycle matching, making the design well-suited for resource-constrained edge-computing platforms.

\subsubsection{Optimized Sliding-Window Duplicate Removal}
\label{subsubsec:duplicate_removal}

The \texttt{Duplicate\_Removal} block filters redundant segments that arise when the same physical line is detected through multiple chain paths. The architecture employs a compact sliding-window register bank that stores the last five unique segments. This approach is effective because duplicates typically occur within a local temporal window, and storing five segments has been shown to be sufficient for practical scenarios.

The filtering process is performed in two stages to minimize switching activity and power consumption:

\begin{enumerate}
    \item \textbf{Angular Pre-Filtering (XOR-based):} For each incoming segment, a lightweight XOR-based comparison is first performed against the angle classes of all stored segments. Candidates with angular deviation exceeding a configurable threshold (\(\theta_{\text{thresh}} = 15^\circ\)) are discarded immediately without proceeding to the geometric comparison stage. This operation significantly reduces switching activity in the arithmetic units.

    \item \textbf{Parallel Geometric Comparison:} The remaining candidates, which have similar orientation, are compared in parallel using dedicated logic that checks both angular difference and spatial distance. A segment is flagged as a duplicate only if it satisfies both conditions:
    \begin{equation}
    |\theta_{\text{in}} - \theta_{\text{stored}}| < \theta_{\text{thresh}}
    \end{equation}
    \begin{equation}
    \text{distance}(\text{seg}_{\text{in}}, \text{seg}_{\text{stored}}) < d_{\text{thresh}}
    \end{equation}
    where \(\theta_{\text{thresh}} = 15^\circ\) and \(d_{\text{thresh}} = 10\) pixels are configurable design parameters.
\end{enumerate}

To illustrate the duplicate removal process, consider the sliding window history shown in Table~\ref{tab:dup_history}. A new segment arrives with:
\[
\text{seg}_{\text{in}} = (x_1=12, y_1=22, x_2=32, y_2=42, \text{angle}=2)
\]

\begin{table}
\centering
\caption{History Buffer Content Before Processing a New Segment}
\label{tab:dup_history}
\begin{tabular}{c|c|c|c|c|c}
\hline
\textbf{Index} & \textbf{x1} & \textbf{y1} & \textbf{x2} & \textbf{y2} & \textbf{Angle} \\
\hline
1 & 10 & 20 & 30 & 40 & 2 \\
2 & 50 & 60 & 70 & 80 & 5 \\
3 & 15 & 25 & 35 & 45 & 2 \\
4 & 100 & 110 & 120 & 130 & 1 \\
5 & 20 & 30 & 40 & 50 & 2 \\
\hline
\end{tabular}
\end{table}

The XOR-based angular pre-filtering compares the new segment's angle (2) with all stored segments. Segments 2 and 4 are discarded, while segments 1, 3, and 5 proceed to geometric comparison.

The geometric comparison results are shown in Table~\ref{tab:dup_comparison}. The distances are computed as Euclidean distances between segment endpoints. For example, for Segment 1, the distance is:
\[
\sqrt{(12-10)^2 + (22-20)^2} = \sqrt{4+4} \approx 2.83
\]

\begin{table}
\centering
\caption{Parallel Geometric Comparison of Candidates. \texttt{Seg.}: Segment; \texttt{Dist.}: Distance; \texttt{Met?}: Condition Met; \texttt{Res.}: Result.}
\label{tab:dup_comparison}
\begin{tabular}{p{0.5cm}|p{2cm}|p{2cm}|p{0.5cm}|p{1.5cm}}
\hline
\textbf{Seg.} & \textbf{Angle Diff. ($^\circ$)} & \textbf{Dist. (pixels)} & \textbf{Met?} & \textbf{Res.} \\
\hline
1 & $|2-2| \times 22 = 0 < 15$ & $2.83 < 10$ & Yes & Dup. \\
3 & $0 < 15$ & $4.24 < 10$ & Yes & Dup. \\
5 & $0 < 15$ & $11.31 \geq 10$ & No & Not Dup. \\
\hline
\end{tabular}
\end{table}

Since the new segment is a duplicate of Segments 1 and 3, it is discarded. If unique, it would replace the oldest entry (Segment 1) to maintain the window size of five.

The use of angular pre-filtering ensures that only geometrically similar segments undergo full comparison, reducing dynamic power consumption by minimizing switching activity in the arithmetic comparison units. The sliding-window approach eliminates the need for large memory buffers while maintaining effective duplicate detection.

\subsubsection{Serializer FIFO}
\label{subsubsec:serializer}

The \texttt{Serializer\_FIFO} throttles the output from the back-end to a steady stream of one segment per clock cycle. The chain builder can generate up to three segments in a single cycle—one for the current horizontal run and up to two for terminated chains—which are then serialized to match the downstream interface.

The block packs the segment information (start and end coordinates, length, and angle class) into standardized output data packets and buffers them in a FIFO queue. The FIFO depth is sized to absorb the variable generation rate of the chain builder while ensuring that no segments are lost during peak output periods. The design uses standard handshake signals (\texttt{out\_valid}, \texttt{out\_ready}) to manage data flow with the downstream system, making it compatible with AXI-Stream or similar interfaces.

The implementation is lightweight, requiring only a small FIFO and minimal control logic, making it directly portable to standard-cell libraries without modification.

\section{Implementation Result}
\label{sec:implementation_result}

\subsection{RTL Implementation}
\label{subsec:rtl_implementation}

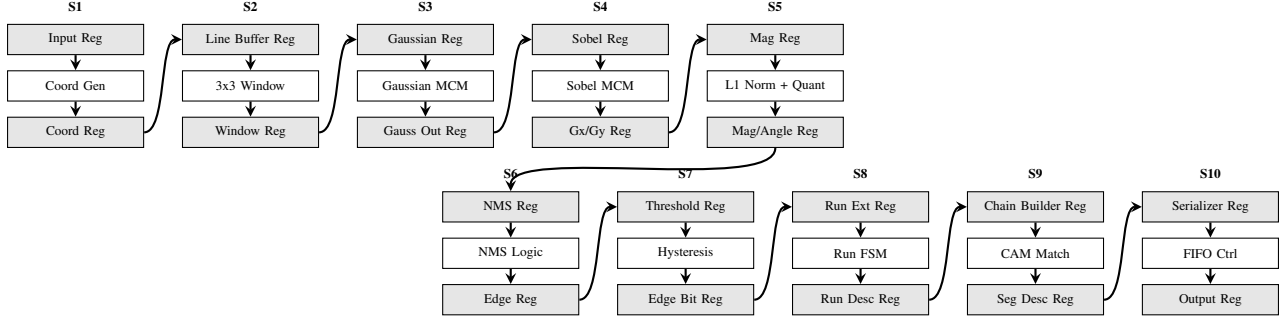
\begin{figure*}
\centering
\begin{tikzpicture}[
    node distance=0.2cm and 0.1cm,
    reg/.style={rectangle, draw, fill=black!10, minimum width=1.8cm, minimum height=0.4cm, align=center, font=\tiny},
    comb/.style={rectangle, draw, fill=white, minimum width=1.8cm, minimum height=0.4cm, align=center, font=\tiny},
    arrow/.style={->, >=stealth, thick}
]


\node[reg] (r1_s1_reg1) {Input Reg};
\node[comb, below=of r1_s1_reg1] (r1_s1_comb1) {Coord Gen};
\node[reg, below=of r1_s1_comb1] (r1_s1_reg2) {Coord Reg};

\node[reg, right=0.5cm of r1_s1_reg1] (r1_s2_reg1) {Line Buffer Reg};
\node[comb, below=of r1_s2_reg1] (r1_s2_comb1) {3x3 Window};
\node[reg, below=of r1_s2_comb1] (r1_s2_reg2) {Window Reg};

\node[reg, right=0.5cm of r1_s2_reg1] (r1_s3_reg1) {Gaussian Reg};
\node[comb, below=of r1_s3_reg1] (r1_s3_comb1) {Gaussian MCM};
\node[reg, below=of r1_s3_comb1] (r1_s3_reg2) {Gauss Out Reg};

\node[reg, right=0.5cm of r1_s3_reg1] (r1_s4_reg1) {Sobel Reg};
\node[comb, below=of r1_s4_reg1] (r1_s4_comb1) {Sobel MCM};
\node[reg, below=of r1_s4_comb1] (r1_s4_reg2) {Gx/Gy Reg};

\node[reg, right=0.5cm of r1_s4_reg1] (r1_s5_reg1) {Mag Reg};
\node[comb, below=of r1_s5_reg1] (r1_s5_comb1) {L1 Norm + Quant};
\node[reg, below=of r1_s5_comb1] (r1_s5_reg2) {Mag/Angle Reg};

\draw[arrow] (r1_s1_reg2.east) .. controls +(0.3cm,0) and +(-0.3cm,0) .. (r1_s2_reg1.west);
\draw[arrow] (r1_s2_reg2.east) .. controls +(0.3cm,0) and +(-0.3cm,0) .. (r1_s3_reg1.west);
\draw[arrow] (r1_s3_reg2.east) .. controls +(0.3cm,0) and +(-0.3cm,0) .. (r1_s4_reg1.west);
\draw[arrow] (r1_s4_reg2.east) .. controls +(0.3cm,0) and +(-0.3cm,0) .. (r1_s5_reg1.west);

\draw[arrow] (r1_s1_reg1) -- (r1_s1_comb1);
\draw[arrow] (r1_s1_comb1) -- (r1_s1_reg2);
\draw[arrow] (r1_s2_reg1) -- (r1_s2_comb1);
\draw[arrow] (r1_s2_comb1) -- (r1_s2_reg2);
\draw[arrow] (r1_s3_reg1) -- (r1_s3_comb1);
\draw[arrow] (r1_s3_comb1) -- (r1_s3_reg2);
\draw[arrow] (r1_s4_reg1) -- (r1_s4_comb1);
\draw[arrow] (r1_s4_comb1) -- (r1_s4_reg2);
\draw[arrow] (r1_s5_reg1) -- (r1_s5_comb1);
\draw[arrow] (r1_s5_comb1) -- (r1_s5_reg2);

\node[above=0.05cm of r1_s1_reg1, font=\tiny\bfseries] {S1};
\node[above=0.05cm of r1_s2_reg1, font=\tiny\bfseries] {S2};
\node[above=0.05cm of r1_s3_reg1, font=\tiny\bfseries] {S3};
\node[above=0.05cm of r1_s4_reg1, font=\tiny\bfseries] {S4};
\node[above=0.05cm of r1_s5_reg1, font=\tiny\bfseries] {S5};


\node[reg, below=1.8cm of r1_s5_reg1, xshift=-3.5cm] (r2_s6_reg1) {NMS Reg};
\node[comb, below=of r2_s6_reg1] (r2_s6_comb1) {NMS Logic};
\node[reg, below=of r2_s6_comb1] (r2_s6_reg2) {Edge Reg};

\node[reg, right=0.5cm of r2_s6_reg1] (r2_s7_reg1) {Threshold Reg};
\node[comb, below=of r2_s7_reg1] (r2_s7_comb1) {Hysteresis};
\node[reg, below=of r2_s7_comb1] (r2_s7_reg2) {Edge Bit Reg};

\node[reg, right=0.5cm of r2_s7_reg1] (r2_s8_reg1) {Run Ext Reg};
\node[comb, below=of r2_s8_reg1] (r2_s8_comb1) {Run FSM};
\node[reg, below=of r2_s8_comb1] (r2_s8_reg2) {Run Desc Reg};

\node[reg, right=0.5cm of r2_s8_reg1] (r2_s9_reg1) {Chain Builder Reg};
\node[comb, below=of r2_s9_reg1] (r2_s9_comb1) {CAM Match};
\node[reg, below=of r2_s9_comb1] (r2_s9_reg2) {Seg Desc Reg};

\node[reg, right=0.5cm of r2_s9_reg1] (r2_s10_reg1) {Serializer Reg};
\node[comb, below=of r2_s10_reg1] (r2_s10_comb1) {FIFO Ctrl};
\node[reg, below=of r2_s10_comb1] (r2_s10_reg2) {Output Reg};

\draw[arrow] (r2_s6_reg2.east) .. controls +(0.3cm,0) and +(-0.3cm,0) .. (r2_s7_reg1.west);
\draw[arrow] (r2_s7_reg2.east) .. controls +(0.3cm,0) and +(-0.3cm,0) .. (r2_s8_reg1.west);
\draw[arrow] (r2_s8_reg2.east) .. controls +(0.3cm,0) and +(-0.3cm,0) .. (r2_s9_reg1.west);
\draw[arrow] (r2_s9_reg2.east) .. controls +(0.3cm,0) and +(-0.3cm,0) .. (r2_s10_reg1.west);

\draw[arrow] (r2_s6_reg1) -- (r2_s6_comb1);
\draw[arrow] (r2_s6_comb1) -- (r2_s6_reg2);
\draw[arrow] (r2_s7_reg1) -- (r2_s7_comb1);
\draw[arrow] (r2_s7_comb1) -- (r2_s7_reg2);
\draw[arrow] (r2_s8_reg1) -- (r2_s8_comb1);
\draw[arrow] (r2_s8_comb1) -- (r2_s8_reg2);
\draw[arrow] (r2_s9_reg1) -- (r2_s9_comb1);
\draw[arrow] (r2_s9_comb1) -- (r2_s9_reg2);
\draw[arrow] (r2_s10_reg1) -- (r2_s10_comb1);
\draw[arrow] (r2_s10_comb1) -- (r2_s10_reg2);

\node[above=0.05cm of r2_s6_reg1, font=\tiny\bfseries] {S6};
\node[above=0.05cm of r2_s7_reg1, font=\tiny\bfseries] {S7};
\node[above=0.05cm of r2_s8_reg1, font=\tiny\bfseries] {S8};
\node[above=0.05cm of r2_s9_reg1, font=\tiny\bfseries] {S9};
\node[above=0.05cm of r2_s10_reg1, font=\tiny\bfseries] {S10};

\draw[arrow] (r1_s5_reg2.south) .. controls +(0,-0.6cm) and +(0,0.6cm) .. (r2_s6_reg1.north);

\end{tikzpicture}
\caption{RTL Datapath: The pipeline comprises ten stages with registered interfaces. The top row (S1--S5) processes pixel input through coordinate generation, line buffering, Gaussian and Sobel filtering, and angle encoding. The bottom row (S6--S10) performs non-maximum suppression, double thresholding, run extraction, CAM-like chain building, and FIFO serialization. Gray blocks are pipeline registers; white blocks are combinational logic.}
\label{fig:rtl_datapath}
\end{figure*}

The proposed architecture is implemented as a fully pipelined RTL datapath with registered interfaces between each processing stage. Fig.~\ref{fig:rtl_datapath} illustrates the pipeline structure, showing the flow of data through ten stages, while Table~\ref{tab:pipeline_stages} summarizes the registers and combinational logic at each stage.

\begin{table*}
\centering
\caption{Pipeline Stages: Registers and Combinational Logic}
\label{tab:pipeline_stages}
\begin{tabular}{c|c|c|c}
\hline
\textbf{Stage} & \textbf{Block} & \textbf{Input Register(s)} & \textbf{Output Register(s)} \\
\hline
1 & Pixel Coord Gen & \texttt{pixel\_in[7:0]}, \texttt{valid} & \texttt{pix\_reg}, \texttt{x\_reg}, \texttt{y\_reg}, \texttt{valid\_reg} \\
2 & Line Buffer & \texttt{pix\_reg} & \texttt{win\_p00..p22[7:0]} (combinational) \\
3 & Gaussian Filter & \texttt{win\_p00..p22} & \texttt{gauss\_out[9:0]} \\
4 & Sobel Core & \texttt{gauss\_out} & \texttt{Gx[9:0]}, \texttt{Gy[9:0]} \\
5 & Mag/Angle Encoder & \texttt{Gx}, \texttt{Gy} & \texttt{mag[7:0]}, \texttt{angle[2:0]} \\
6 & Modified NMS & \texttt{mag}, \texttt{angle} & \texttt{edge\_candidate[0:0]} \\
7 & Double Threshold & \texttt{mag}, \texttt{edge\_candidate} & \texttt{edge\_bit[0:0]} \\
8 & Run Extractor & \texttt{edge\_bit} & \texttt{run\_desc} \\
9 & Chain Builder & \texttt{run\_desc} & \texttt{seg\_desc} \\
10 & Serializer FIFO & \texttt{seg\_desc} & \texttt{output\_packet} \\
\hline
\end{tabular}
\end{table*}

The total pipeline delay is \(5W + 27\) clock cycles, where \(W\) is the image width, as derived in~\cite{Ossimitz2021FPGA}. The register-based design ensures deterministic latency and eliminates the need for memory controllers, making it well-suited for real-time edge-computing applications.

\subsection{Functional Verification}
\label{subsec:functional_verification}

\begin{figure*}
    \centering
    \includegraphics[width=1.5\columnwidth]{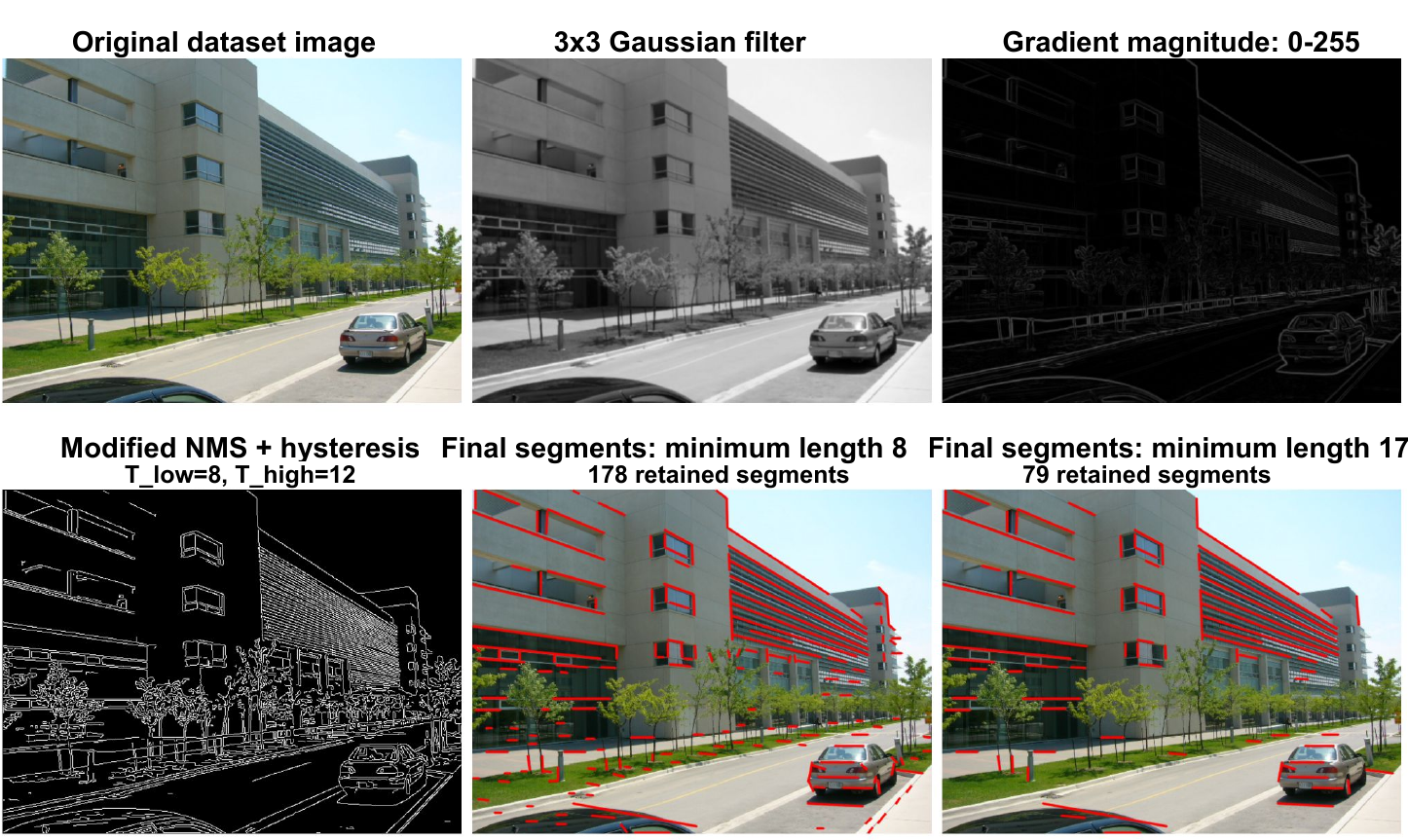}
    \caption{Functional verification of the proposed pipeline on a YorkUrban dataset image. Top row: original input, \(3\times3\) Gaussian-smoothed image, and L1-norm gradient magnitude. Bottom row: binary edge map after modified NMS and hysteresis thresholding (\(T_{\text{low}}=8\), \(T_{\text{high}}=12\)), and final detected segments overlaid in red for minimum segment lengths of 8 (178 segments) and 17 (79 segments).}
    \label{fig:functional_validation}
\vspace{-1em}
\end{figure*}

Before hardware synthesis, the complete detection pipeline was validated using a bit-accurate software model executed on images from the YorkUrban dataset~\cite{yorkurban}. Fig. ~\ref{fig:functional_validation} illustrates the intermediate and final outputs of each processing stage for a representative urban scene. The \(3\times3\) Gaussian filter (Eq.~\ref{eq:gaussian_mcm}) suppresses sensor noise while preserving the strong structural edges of the building facade. The resulting L1-norm gradient magnitude map (Eq.~\ref{eq:magnitude_l1}) highlights architectural contours clearly against a low-response background. The modified non-maximum suppression, combined with hysteresis thresholding (\(T_{\text{low}}=8\), \(T_{\text{high}}=12\)), produces a single-pixel-wide edge map with continuous contours along window rows, balcony rails, and road boundaries. This confirms the edge-continuity benefit of checking neighboring pixels against their own orthogonal neighbors, as described in Section~\ref{subsubsec:modified_nms}. The final two panels show the segments extracted by the run extractor, CAM-like chain builder, and duplicate removal back-end under two minimum-length settings. A threshold of 8 pixels retains 178 segments, capturing fine structural detail including short window frames and curb markings. Raising the threshold to 17 pixels retains 79 segments that correspond to the dominant structural lines of the scene. This 56\% reduction in segment count, with negligible loss of salient structure, demonstrates that the minimum-length parameter provides an effective means of trading detection density against downstream processing load. It also confirms that the fixed 64-entry chain capacity and five-entry duplicate-removal window are sufficient for real-world urban imagery. The number of concurrently active chains never exceeded the 64-entry capacity during these runs, which is consistent with the dataset analysis in Section~\ref{subsubsec:chain_builder}.

\subsection{Synthesis Results}
\label{subsec:synthesis_results}

The proposed ASIC architecture was described in RTL-level VHDL and synthesized using Synopsys Design Compiler (version L-2016.03-SP1) with the FreePDK45 standard-cell library, which corresponds to a 45nm CMOS technology node. Synthesis was performed at two operating frequencies: 100 MHz as the nominal frequency and 125 MHz to evaluate the maximum achievable performance. All synthesis runs were performed under typical operating conditions.

The synthesized design consists of 126,220 standard cells, comprising 109,660 combinational cells and 16,497 sequential cells (flip-flops). The total cell area is 412,181 \textmu m\textsuperscript{2} (0.412 mm\textsuperscript{2}), with combinational logic contributing 282,058 \textmu m\textsuperscript{2} and sequential logic contributing 130,123 \textmu m\textsuperscript{2}. Table~\ref{tab:area_power} summarizes the area breakdown and power consumption at both operating frequencies.

\begin{table}
\centering
\caption{Area Breakdown and Power Consumption Summary. Area in \textmu m\textsuperscript{2}; Power in $mW$.}
\label{tab:area_power}
\begin{tabular}{l|c|c|c}
\hline
\textbf{Component} & \textbf{Area} & \textbf{100 MHz} & \textbf{125 MHz} \\
\hline
Combinational Cells & 282,058 & 0.83 & 1.03 \\
Sequential Cells (Flip-Flops) & 130,123 & 22.95 & 28.69 \\
\hline
\textbf{Total} & \textbf{412,181} & \textbf{25.54} & \textbf{31.48} \\
\hline
\end{tabular}
\end{table}

The dominance of register power is expected for a heavily pipelined design with 16,497 flip-flops. At 125 MHz, the dynamic power scales linearly with frequency while leakage power remains unchanged.

\subsection{Timing and Performance Discussion}
\label{subsec:timing_performance}

The critical path is located in the run extractor module, specifically in the path from \texttt{run\_count\_reg} to \texttt{run\_angle\_hist\_reg}, which involves a 10-bit multiplication followed by a division operation. At 100 MHz, the worst-case slack is 1.98 ns, indicating a comfortable timing margin. At 125 MHz, the slack reduces to 0.01 ns, confirming that this is the maximum operating frequency for the design.

Table~\ref{tab:performance} summarizes the performance metrics for different resolutions at 100 MHz.

\begin{table}
\centering
\caption{Performance Summary at 100 MHz}
\label{tab:performance}
\begin{tabular}{l|c|c|c}
\hline
\textbf{Resolution} & \textbf{Width (W)} & \textbf{Latency (\textmu s)} & \textbf{FPS} \\
\hline
VGA (640\texttimes480) & 640 & 32.27 & 325 \\
HD (1280\texttimes720) & 1280 & 64.27 & 108 \\
Full HD (1920\texttimes1080) & 1920 & 96.27 & 48 \\
\hline
\end{tabular}
\end{table}

The synthesis results validate the effectiveness of the three key ASIC optimizations introduced in this work. The register-based line buffer with data reuse reduces net switching power to only 0.69 mW, which corresponds to 2.7\% of total power, confirming that the circular pointer mechanism successfully minimizes data movement. The multiplierless MCM-based filtering eliminates all 27 multipliers per pixel, reducing combinational area to 0.282 mm\textsuperscript{2}, which is substantially smaller than designs relying on CORDIC or DSP-based multipliers. The CAM-like chain builder enables deterministic single-cycle matching, achieving a latency of 32.27 \textmu s at 100 MHz. This is over 150 times faster than the 5 ms latency reported in existing ASIC implementations~\cite{Pachkor2018Memory}. The fixed capacity of 64 chains also reduces storage requirements, contributing to the compact sequential area of 0.130 mm\textsuperscript{2}.

Table~\ref{tab:comparison} compares the proposed ASIC architecture with existing hardware implementations. A direct area comparison is not possible, as the prior ASIC works~\cite{Majumdar2000ASIC,Pachkor2018Memory} do not report explicit silicon area figures. Furthermore, these designs are based on the Hough Transform, which inherently requires large parameter-space memories, whereas the step-length implementation presented here uses a compact CAM-like structure. Despite this limitation, the proposed design achieves the lowest power consumption among all implementations (25.54 mW) while delivering the highest frame rate among ASIC-based works.

\begin{table}[]
\centering
\caption{Comparison with Prior Hardware Implementations. Area is not reported in \cite{Majumdar2000ASIC} and \cite{Pachkor2018Memory}.}
\label{tab:comparison}
\begin{tabular}{l|c|c|c}
\hline
\textbf{Work} & \textbf{Tech.} & \textbf{Power (mW)} & \textbf{FPS} \\
\hline
Ossimitz \cite{Ossimitz2021FPGA} & 28nm (FPGA) & 64 & 325 \\
Majumdar \cite{Majumdar2000ASIC} & 180nm (ASIC) & \(\sim\)100 & \(\sim\)20 \\
Pachkor \cite{Pachkor2018Memory} & 90nm (ASIC) & \(\sim\)50 & \(\sim\)200 \\
\hline
\textbf{This Work (100 MHz)} & \textbf{45nm (ASIC)} & \textbf{25.54} & \textbf{325} \\
\textbf{This Work (125 MHz)} & \textbf{45nm (ASIC)} & \textbf{31.48} & \textbf{406} \\
\hline
\end{tabular}
\end{table}

The combination of these optimizations results in a total power consumption of 25.54 mW at 100 MHz, which is 60\% lower than the FPGA implementation (64 mW), 74\% lower than the 180nm ASIC (100 mW), and 49\% lower than the 90nm ASIC (50 mW). At 125 MHz, the total power increases to 31.48 mW (23\% increase) while delivering 406 FPS, demonstrating the scalability of the design.

\section{Conclusion}
\label{sec:conclusion}

This paper has presented a low-latency ASIC architecture for real-time line segment detection, targeting edge-computing applications where deterministic timing, low power consumption, and minimal silicon area are critical. The proposed design is based on the step-length algorithm and incorporates five ASIC-specific features: register-based line buffering with data reuse, multiplierless MCM-based filtering, 8-class angle quantization, CAM-like associative memory for single-cycle matching, and an optimized duplicate removal mechanism.

The synthesis results demonstrate that these optimizations collectively achieve a total cell area of 0.412 mm\textsuperscript{2} and power consumption of 25.54 mW at 100 MHz, with 325 FPS and 32.27 \textmu s latency at VGA resolution. The multiplierless MCM approach eliminates all 27 multipliers per pixel, reducing combinational area to 0.282 mm\textsuperscript{2}. The data reuse technique minimizes net switching power to 0.69 mW, which corresponds to 2.7\% of total power. The CAM-like chain builder enables over 150 times faster latency than prior ASIC implementations. Compared with existing designs, the proposed architecture achieves the lowest reported power consumption: 60\% lower than the FPGA implementation (64 mW), 74\% lower than the 180nm ASIC (100 mW), and 49\% lower than the 90nm ASIC (50 mW).

The fully pipelined design with content-independent latency makes it well suited for real-time embedded vision systems in autonomous driving, drone navigation, and industrial inspection. Future work includes exploring advanced clock gating techniques, extending the architecture to support 4K resolution, and integrating a lightweight neural network for semantic line detection in complex scenes.

\bibliographystyle{IEEEtran}
\bibliography{References}

\end{document}